\documentclass[12pt,english]{article}
\usepackage[T1]{fontenc}
\usepackage[latin1]{inputenc}
\usepackage{amsmath}
\usepackage{amssymb}
\makeatletter

\catcode`@=12
\usepackage{babel}
\makeatother
\begin{document}
\thispagestyle{empty}

\mbox{} \vspace{0.75in}

\begin{center}\textbf{\large CP violation in neutrino mass matrix}\\
 \vspace{1.0in} \textbf{Utpal Sarkar and Santosh K. Singh}\\
 \vspace{0.2in} \textsl{${}^{2}$ Physical Research Laboratory, Ahmedabad
380009, India}\\
 \vspace{.75in}\end{center}

\begin{abstract}
\

We constructed rephasing invariant measures of CP violation with elements
of the neutrino mass matrix, in the basis in which the charged lepton
mass matrix is diagonal. We discuss some examples of neutrino mass
matrices with texture zeroes, where the present approach is applicable
and demonstrate how it simplifies an analysis of CP violation.
We applied our approach to study
CP violation in all the phenomenologically acceptable
3-generation two-zero texture neutrino mass matrices and shown that 
in any of these cases there is only one CP phase which contributes
to the neutrino oscillation experiment and there are no Majorana phases. 

\end{abstract}
\newpage
\baselineskip 20pt

\section{Introduction}

In the standard model there is only one source of CP violation, which
is in the charged-current mixing matrix in the quark sector. 
The charged-current
mixing matrix in the quark sector contains one CP phase, which has
been observed. It is not possible to identify the position of the
CP phase, since it is possible to make any phase transformations to
the quarks. However, it is possible to define a rephasing invariant
quantity as product of elements of the mixing matrix that remains
invariant under any rephasing of the quarks \cite{cp1,cp2}. This
is known as Jarlskog invariant.

In the leptonic sector, standard model does not allow any CP violation.
If one considers extensions of the standard model to accommodate the
observed neutrino masses, then there can be several CP phases 
\cite{cp3,cp4,cp5,cp6}.
In the simplest scenario of three generations, there could be one
CP phase in the mixing matrix in the leptonic sector, similar to the
quark sector. In addition, if neutrinos are Majorana particles they
can have two more Majorana CP phases \cite{cp4}. In this case it
is possible to work in a parametrization, in which all the three CP
phases could be in the charged-current mixing matrix in the leptonic
sector. One of these CP phase will contribute to the neutrino oscillation
experiments, while the other two will contribute to lepton number
violating process like neutrinoless double beta decay. A natural explanation
for the smallness of the neutrino masses comes from the see-saw mechanism
\cite{seesaw}. The origin of small neutrino mass then relates to
a large lepton number violating scale. It is quite natural that this
lepton number violation at the high scale would also explain the baryon
asymmetry of the universe through leptogenesis \cite{lepto1,lepto2}.
This connection between the neutrino mass and leptogenesis makes
the question of CP violation in the leptonic sector more interesting
\cite{cp5,cp6}.

The CP phases in the leptonic sector has been studied and rephasing
invariants for both lepton number conserving as well as lepton number
violating CP violation have been constructed \cite{cp3}. In this
article we try to study this question only in terms of neutrino masses.
Since neutrinos are produced only through weak interactions, it is
possible to work in the weak interaction basis, in which the charged
lepton mass matrix is diagonal. The neutrino mass matrix in this basis
will then contain all the information about CP violation. We try to
find rephasing invariant combinations of the neutrino mass elements,
so that with those invariants some general comments can be made about
CP violation in the model without deriving the structure of the charged-current
mixing matrix.

\section{CP Violation in the Quark Sector}

We briefly review the rephasing invariants in terms of the mixing
matrices and then show how the same results can be obtained from the
mass matrix without taking the trouble of diagonalizing them in the
leptonic sector. Consider first the quark sector, where the up and
the down quark mass matrices are diagonalized by the bi-unitary transformations.
We write the corresponding unitary matrices that relates the left-handed
and right-handed physical (with definite masses) up and down quarks
fields to their weak (diagonal charged current) basis as: $U_{L},D_{L},U_{R}$
and $D_{R}$. Then the charged current interactions in terms of the
physical fields will contain the Kobayashi-Cabibbo-Maskawa mixing
matrix \[
V=D_{L}^{\dagger}U_{L}.\]
 Since the right-handed fields are singlets under the standard model
interactions, they do not enter in the charged current interactions.
In any physical processes, only this CKM mixing matrix would appear
and hence the matrices $U_{R}$ and $D_{R}$ becomes redundant. So,
the up and down quark masses have much more freedom and the physical
observables that can determine the $V_{\alpha i}$ cannot infer about
the up and down quark masses uniquely.

For the CP violation, one needs to further consider the rephasing
of the left-handed fields. Any phase transformation to the up and
down quarks will also transform the CKM matrix \[
V_{\alpha i}\rightarrow e^{-i(d_{\alpha}-u_{i})}.\]
 However, if there is any CP phase in the CKM matrix, which cannot
be removed by any phase transformations of the up and the down quarks,
should be present in the Jarlskog invariant \cite{cp1,cp2}\begin{equation}
J_{\alpha i\beta j}={\textrm{Im}}[V_{\alpha i}V_{\beta j}V_{\alpha j}^{\ast}V_{\beta i}^{\ast}].\end{equation}
 Thus if the Jarlskog invariant is non-vanishing, that would imply
CP violation in the quark mixing. It is apparent from the definition
that any phase transformations to the up and down quarks cannot change
$J_{\alpha i\beta j}$. In a three generation scenario there can be
only one such invariant and hence the CKM matrix can have only one
CP phase, which is invariant under rephasing of the up and the down
quarks.

\section{CP Violation in the Leptonic Sector}

In the leptonic sector, the charged lepton mass matrix can in general
be diagonalized by a bi-unitary transformation. In addition, the neutrinos
are produced in weak interactions, so the flavour of the neutrinos
at the time of production is always same as the flavour of the charged
lepton. The charged-current interaction is given by \begin{equation}
{\mathcal{L}}_{cc}={\frac{g}{\sqrt{2}}}~\bar{\ell}_{iL}\gamma^{\mu}\nu_{L}~W_{\mu}^{-}\end{equation}
 in the basis ($\ell_{iL},i=e,\mu,\tau$) in which charged lepton
mass matrix is diagonal, \textit{i.e.}, the states $e,\mu,\tau$ correspond
to physical states. Without loss of generality we further assume that
elements of the diagonal charged lepton mass matrix are real and positive.

The neutrino mass term can be written as \begin{equation}
{\mathcal{L}}_{M}=M_{\nu ij}~\overline{{\nu^{c}}_{iR}}~\nu_{jL}=M_{\nu ij}~\nu_{iL}^{T}~C^{-1}~\nu_{jL}\end{equation}
 so that the neutrino mass matrix can be diagonalized by a single
unitary matrix $U$ through \begin{equation}
U^{T}~M_{\nu}~U=K^{2}~\hat{M}_{\nu}\label{nu-diag}\end{equation}
 where $\hat{M}_{\nu}={\textrm{diag}}[m_{1},~m_{2},~m_{3}]$ is a
real diagonal matrix and $K$ is a diagonal phase matrix. The unitary
matrix $U_{ia}$ (with $i,j=e,\mu,\tau$ and $a,b=1,2,3$) relates
the physical neutrino states $\nu_{a}$ (with masses $m_{a}$) to
the weak states \begin{equation}
\nu_{a}=U_{ai}^{\ast}~\nu_{i}+{K_{aa}^{\ast}}^{2}~U_{ai}~\nu_{i}^{c},\end{equation}
 so that the physical neutrinos satisfy the Majorana condition \begin{equation}
\nu=K^{2}~\nu^{c}.\end{equation}
 The Unitary matrix $U$ thus gives the mixing of the neutrinos and
hence neutrino oscillations, which is known as the PMNS mixing matrix
\cite{pmns} and $K$ is the Majorana phase matrix containing the
Majorana phases, which are the new sources of CP violation entering
due to the Majorana nature of the neutrinos. The PMNS matrix $U$
also can contain CP violating phases, which should be observed in
the neutrino oscillation experiments. We call these phases in the
PMNS mixing matrix $U$ as {\it Dirac phases} to distinguish them from the 
{\it Majorana phases}. The main difference betwen a Majorana phase
and a Dirac phase is that the Majorana phases do not affect any
lepton number conserving process like neutrino oscillations. On the
other hand, the Dirac phases may contribute to both lepton number
conserving as well as lepton number violating processes. 

From the above discussions it is apparent that the information about
the CP phases can be obtained from either $U$ and $K$ or only from
the mass matrix $M_{\nu}$. In the literature the question of CP violation
is usually discussed by studying $U$ and $K$. In this article we
point out that it is possible to study the question of CP phases only
by studying the neutrino mass matrix $M_{\nu}$. In particular, the
information about CP violation is conveniently obtained from the rephasing
invariant combinations of neutrino mass elements. When the neutrino
masses originate from see-saw mechanism, the question of CP violation
has been studied in details and similar invariants have been constructed
\cite{cp6}. Our approach is different and we do not restrict our
analysis to any specific origin of the neutrino masses. Our results
are general and applicable to any models of neutrino masses.

Consider the transformation of different quantities under the rephasing
of the neutrinos \begin{eqnarray}
\nu_{i} & \rightarrow & e^{i\delta_{i}}\nu_{i}\nonumber \\
\ell_{a} & \rightarrow & e^{i\eta_{a}}\ell_{a}\nonumber \\
U_{ai} & \rightarrow & e^{-i(\eta_{a}-\delta_{i})}U_{ai}\nonumber \\
K_{i} & \rightarrow & e^{i\delta_{i}}K_{i}.\end{eqnarray}
 From these transformations it is possible to construct the rephasing
invariants \cite{cp3}\begin{equation}
s_{aij}=U_{ai}U_{aj}^{*}K_{i}^{*}K_{j}.\end{equation}
 In the three generation case there will be three independent rephasing
invariant measures. There is another rephasing invariant which is
similar to the Jarlskog invariant in the quark sector, \begin{equation}
t_{aibj}=U_{ai}U_{bj}U_{aj}^{\ast}U_{bi}^{\ast},\end{equation}
 so that $T_{aibj}={\textrm{Im}}~t_{aibj}$ and $S_{aij}={\textrm{Im}}~s_{aij}$
becomes the measure of CP violation. $T_{aibj}$ contains the information
about the Dirac phase, while $s_{aji}$ contains information about 
both Dirac as well as Majorana phases. 
One can then use the relation
\[ t_{aibj}=s_{aij}\cdot s_{aji}\]
to eliminate the invariants $T$'s or else keep the $T$'s as independent
measures and reduce the number of independent $S$'s. One convenient
choice for the independent measures is the independent $t_{aibj}$s
and $s_{1ij}$s. In the three generation case there is only one $t_{aibj}$
and two $s_{1ij}$s. The advantage of this parametrization is that
the measure $T_{aibj}$ gives the CP violation in any neutrino oscillation
experiment, while the measures $S_{1ij}$ corresponds to CP violation
in lepton number violating interactions like the neutrinoless double
beta decay or scattering processes like $W^{-}+W^{-}\rightarrow\ell_{i}^{-}+\ell_{j}^{-}$ also. 

\section{Rephasing Invariants with Neutrino Masses}

We shall now proceed to construct such measures of CP violation in
terms of the mass matrix itself. The rephasing invariant measures
with the mixing matrix can allow all the rephasing invariants
non-vanishing even when there  is only one Dirac phase. However,
in the present formalism, the number of rephasing invariants is
same as the number of CP phases. So, we can find out if there is
any Majorana phase or not. Since the neutrino mass matrix is
diagonalized by a single unitary matrix, the mass matrix contains
all the information about the PMNS mixing matrix and also the mass
eigenstates. However, this is not obvious with the CP phase. When
the neutrinos are given a phase transformation, the mass matrix will
be transformed the same way. Since the we are working in the weak
basis, any transformation to the charged leptons can be transformed
to the mixing matrix and in turn to the neutrino masses. Thus the
phase transformation to the mass matrix will become \begin{eqnarray}
\nu_{i} & \rightarrow & e^{i\delta_{i}}\nu_{i}\nonumber \\
\ell_{i} & \rightarrow & e^{i\eta_{i}}\ell_{i}\nonumber \\
M_{\nu ij} & \rightarrow & e^{i(\delta_{i}+\delta_{j}-\eta_{i}-\eta_{j})}M_{\nu ij}.\end{eqnarray}
 Consider the transformation $E\rightarrow XE$, where X is the phase
transformation to the charged leptons. The mixing matrix will transform
as $U\rightarrow X^{*}U$. However, in equation \ref{nu-diag} this
transformation can be interpreted as a transformation to the mass
matrix, $M\rightarrow X^{*}MX^{*}$. Thus any rephasing invariant
measure constructed with only the mass matrix will contain the information
about CP violation.

Unlike the mixing matrices, the mass matrix is not unitary and instead
it is symmetric. We write the elements of the mass matrix $M_{\nu}$
as $m_{ij}$ and try to construct the rephasing invariants in terms
of $m_{ij}$. This analysis do not depend on the origin of neutrino
masses. We work with the neutrino mass matrix after integrating out
any heavier degrees of freedom and in the weak basis. Any quadratic
terms that can be constructed from the elements of the neutrino mass
matrix are all real, $m_{ij}^{*}m_{ij}=|m_{ij}|^{2}$, as expected.
Let us next consider the quartic terms \begin{equation}
{\mathcal{I}}_{ijkl}=m_{ij}m_{kl}m_{il}^{*}m_{kj}^{*}.\end{equation}
 It is easy to check that any three factors of the above quartic invariant
can be made real by appropriate rephasing, but fourth one will remain
complex. Since there are $n$ re-phasing phases $\left(\delta_{i}\right)$,
one can get $n$ number of linear equations to make mass elements
of the mass matrix to be real. So $n$ number of entries (excluding
symmetric elements) of the mass matrix can be made real, but positions
of the mass entries can not be chosen randomly. That is the reason
why all the above rephasing quartic invariants can not be made real
in general. An $n\times n$ symmetric matrix has $n(n+1)/2$ independent
entries and so it has the same number of phases. By appropriate rephasing,
as argued above, $n$ independent phases can be removed. Then, one
is left with $n(n-1)/2$ number of independent phases.

To find out the minimal set of rephasing invariants we list some of
the transitive and conjugation properties of the invariants: \begin{eqnarray*}
{\mathcal{I}}_{ijpl}{\mathcal{I}}_{pjkl} & = & 
|m_{pj}m_{pl}|^{2}{\mathcal{I}}_{ijkl}\\
{\mathcal{I}}_{ijkp}{\mathcal{I}}_{ipkl} & = & 
|m_{ip}m_{kp}|^{2}{\mathcal{I}}_{ijkl}\end{eqnarray*}
 and \begin{equation}
{\mathcal{I}}_{ijkl}={\mathcal{I}}_{ilkj}^{*}=
{\mathcal{I}}_{klij}={\mathcal{I}}_{kjil}^{*}\label{properties}
\end{equation}
 Using these relations it can be shown that all the ${\mathcal{I}}_{ijkl}$
are not independent and they can be expressed in terms of a subset
of these invariants ${\mathcal{I}}_{ij\alpha\alpha}$ and the quadratic
invariants as \begin{equation}
{\mathcal{I}}_{ijkl}=
\frac{{\mathcal{I}}_{ij\alpha\alpha}{\mathcal{I}}_{kl\alpha\alpha}
{\mathcal{I}}_{li\alpha\alpha}^{*}{\mathcal{I}}_{kj\alpha\alpha}^{*}}
{{|m_{\alpha\alpha}|^{4}|m_{i\alpha}m_{j\alpha}m_{k\alpha}
m_{l\alpha}|^{2}}}\label{quarticinvariant}\end{equation}
 Where $i,j\neq\alpha$ and $\alpha=1,2,...,n$, where $n$ is the
number of generations. On the other hand, any quartics of the form
${\mathcal{I}}_{ij\alpha\alpha}$ can be expressed in terms of 
${\mathcal{I}}_{\beta\beta\alpha\alpha}$
as \begin{equation}
{\textrm{Im}}~[{\mathcal{I}}_{ij\alpha\alpha}]=-{\textrm{Im}}~
[{\mathcal{I}}_{i\alpha\alpha j}]=-{\frac{{\textrm{Im}}
[{\mathcal{I}}_{ii\alpha\alpha}\cdot{\mathcal{I}}_{\alpha\alpha jj}\cdot
{\mathcal{I}}_{iijj}]}{{\textrm{Re}}~
[{\mathcal{I}}_{i\alpha\alpha j}]~(|m_{ii}|^{2}~|m_{jj}|^{2})}}.
\label{cond}
\end{equation}
 Thus we can express all other invariants in terms of ${\mathcal{I}}_{iijj}$
and hence consider them to be of fundamental importance. However,
when there are texture zeroes in the neutrino mass matrix, some or
all of these invariants ${\mathcal{I}}_{iijj}$ could be vanishing.
In that case, it is convenient to use the ${\mathcal{I}}_{ij\alpha\alpha}$
as the measure of CP violation. For the present we shall concentrate
on the more general case with neutrino mass matrices without any texture
zeroes, when the simplest rephasing invariants are ${\mathcal{I}}_{iijj}$.

We can thus define the independent CP violating measures as 
\begin{equation}
I_{ij}={\textrm{Im}}~[{\mathcal{I}}_{iijj}]=
{\textrm{Im}}~[{\mathcal{I}}_{iijj}]=
{\textrm{Im}}~[m_{ii}m_{jj}m_{ij}^{*}m_{ji}^{*}],~~~~~~~~(i<j)
\end{equation}
 These are the minimal set of CP violating measures one can construct
and this gives the independent CP violating quantities. Since $I_{ij}$
satisfies \[
I_{ij}=I_{ji}~~~~~{\textrm{and}}~~~~~I_{ii}=0,\]
there are $n(n-1)/2$ independent measures for $n$ generations.

We ellaborate with some examples starting with a 2-generation scenario.
There are three ${\mathcal{I}}_{ijkl}$, two of which are real: 
$~~{\mathcal{I}}_{1211}=|m_{11}m_{12}|^{2};$
and $~~{\mathcal{I}}_{1222}=|m_{12}m_{22}|^{2}$. The third one can
have imaginary phase, which is 
$I_{12}={\textrm{Im}}[{\mathcal{I}}_{1122}]=
{\textrm{Im }}[m_{11}m_{22}m_{12}^{*}m_{21}^{*}].$
In the 3-generation case there are thus three independent measures
$I_{12},I_{13},I_{23}$. Imaginary phases in all other quartics 
${\mathcal{I}}_{ijkl}$
are related to only these three independent measures. For example,
\[
{\mathcal{I}}_{1223}^{2}= \frac{ I_{12}^* \cdot I_{23}^* \cdot
I_{13} }{|m_{11}|^{2}~|m_{33}|^{2}}.\]
 Similarly, for 4-generations there will be six rephasing invariant
independent phases, which are $I_{12},I_{23},I_{31},I_{14},I_{24},I_{34}$.

The above arguments have been stated without considering any texture
zeroes in the mass matrix. If any element of the mass matrix is zero,
then these discussions have to be generalized. It is because some
quartic invariants can become undefined because of vanishing denominator
of the right hand side of the expression \ref{quarticinvariant} and
\ref{cond}. In that case one needs to consider all possible invariants
${\mathcal{I}}_{ijkl}$, which could be non-vanishing. In addition,
even if all the quartic invariants vanish, the product of six mass
matrix elements of the form \[
{\mathcal{I}}_{ijklpq}=m_{ij}~m_{kl}~m_{pq}~m_{il}^{*}~m_{kq}^{*}~m_{pj}^{*}\]
 could be non-vanishing and can contribute to CP violation. When there
are no texture zeroes, the product of six mass elements do not contain
any new information about CP phases, they are related to the quartic
invariants \begin{equation}
{\mathcal{I}}_{ijklpq}={\frac{{\mathcal{I}}_{ijkl}~
{\mathcal{I}}_{pqkj}}{|m_{kj}|^{2}}}.\end{equation}
 Other products of six mass elements are of the form, 
$m_{ij}~m_{kl}~m_{pq}~m_{il}^{*}~m_{kj}^{*}~m_{pq}^{*}=|m_{pq}|^{2}~
{\mathcal{I}}_{ijkl}$
or $|m_{ij}~m_{kl}~m_{pq}|^{2}$.

We summarize this section by restating that when all elements of the
neutrino mass matrix are non-vanishing, $I_{ij}, ~(i<j)$ gives the
total number of Dirac and Majorana phases. If some of the elements
of the mass matrix vanishes, then either ${\cal I}_{ijkl}$ or
${\cal I}_{ijklpq}$ could also represent some of the independent phases.

\section{CP Violation in Lepton Number Conserving Processes}

The rephasing invariant independent phases contained in $I_{ij},i<j$,
are inclusive of the Dirac phases as well as the
Majorana phases. We shall now identify the rephasing
invariant measures, which is independent of the Majorana phases, which
would enter in the neutrino oscillation experiments. The mass matrix
($M_{\nu}$) in terms of the diagonal mass matrix ($\hat{M}_{\nu}$)
can be expressed following equation \ref{nu-diag} as \[
M_{\nu}=U^{*}~K^{2}~\hat{M}_{\nu}~U^{\dagger}.\]
 Thus the products \begin{equation}
\tilde{M}=(M_{\nu}^{\dagger}~M_{\nu})=(M_{\nu}~M_{\nu}^{\dagger})^{*}=U~\hat{M}_{\nu}^{2}~U^{\dagger}\end{equation}
 are independent of the Majorana phases $K$ and any rephasing invariant
measure constructed with elements $\tilde{m}_{ij}$ of $\tilde{M}$
will contain only the Dirac phases and hence should contribute to any
lepton number conserving processes. 

The mass-squared elements $\tilde{m}_{ij}$ transforms under rephasing
of the neutrinos and charged leptons as \begin{equation}
\tilde{m}_{ij}\rightarrow e^{i(\delta_{i}-\delta_{j})}\tilde{m}_{ij}.\label{mtrephase}\end{equation}
 Neutrino rephasing does not appear because it cancels in $\tilde{M}$.
Since the mass-squared matrix $\tilde{M}_{\nu}$ is Hermitian, $\tilde{M}_{\nu}^{\dagger}=\tilde{M}_{\nu}$,
the mass elements satisfy \begin{equation}
\tilde{m}_{ij}=\tilde{m}_{ji}^{*}.\label{mti}\end{equation}
 Thus the simplest rephasing invariant that can be constructed from
the mass-squared matrix $\tilde{M}_{\nu}$ is just $\tilde{m}_{11}$.
However, from equation \ref{mti} it is obvious that this is a real
quantity. The next possible rephasing invariant would be a quadratic
term, but even that is also real \[
\tilde{m}_{ij}\tilde{m}_{ji}=\tilde{m}_{ij}\tilde{m}_{ij}^{*}=|\tilde{m}_{ij}|^{2}.\]
 Thus the simplest rephasing invariant combination that can contain
the complex CP phase is of the form \begin{equation}
{\mathcal{J}}_{ijk}=\tilde{m}_{ij}~\tilde{m}_{jk}~\tilde{m}_{ki}~~~~~~(i\neq j\neq k).\label{mtinvariant}\end{equation}
 ${\textrm{Im}}[{\mathcal{J}}_{ijk}]$ are antisymmetric under interchange
of any two indices and hence vanishes when any two of the indices
are same. We can express ${\mathcal{J}}_{ijk}$ in terms of $M$ matrix
elements as, \begin{eqnarray}
{\mathcal{J}}_{ijk} & = & \tilde{m}_{ij}\tilde{m}_{jk}\tilde{m}_{kl}\nonumber \\
 & = & \left(\sum_{\alpha}m_{i\alpha}^{*}m_{j\alpha}\right)\left(\sum_{\beta}m_{j\beta}^{*}m_{k\beta}\right)\left(\sum_{\gamma}m_{k\gamma}^{*}m_{l\gamma}\right)\label{vectorproduct}\end{eqnarray}
 Where $\sum_{\alpha}m_{i\alpha}^{*}m_{j\alpha}$ can be interpreted
as scalar product of $i$th and $j$th row. 
A similar invariant was constructed in the case of see-saw
model of neutrino masses in ref. \cite{cp6}, although the
approach to the problem is completely different. In this
expression, if any one
scalar product vanishes then number of independent rephasing measure
${\textrm{Im}}[{\mathcal{J}}_{ijk}]$ which are independent of the
Majorana phases will be reduced by one.

It is possible to express all the rephasing invariants 
containing the Dirac phases ${\mathcal{J}}_{ijk}$
in terms of a minimal set of $\frac{(n-1)(n-2)}{2}$ invariants 
${\mathcal{J}}_{ijn},~(i<j<n)$
as \begin{equation}
{\mathcal{J}}_{ijk}=\frac{{\mathcal{J}}_{ijn}{\mathcal{J}}_{jkn}
{\mathcal{J}}_{kin}}{|m_{in}|~|m_{jn}|~|m_{kn}|}\label{cpms}
\end{equation}
 where $n$ is the index corresponding to the number of generations.
Thus we define the measures of CP violation in lepton number conserving
processes as
\begin{equation}
J_{ijn} = {\rm Im}[{\cal J}_{ijn}]  ~~~~~~~~~~~(i < j < n).
\end{equation}
These invariants ${\textrm{Im}}[{\mathcal{J}}_{ijk}]$ are not independent
of the invariants ${\mathcal{I}}_{ijkl}$ and can be exressed as 
\begin{equation}
{\mathcal{J}}_{ijk}=\sum_{a,b,c}{\frac{{\mathcal{I}}_{iajb}\cdot
{\mathcal{I}}_{kaic}}{|m_{ia}|^{2}}}.\end{equation}
So, the independent measures $I_{ij}$ include these independent
measures of Dirac CP phases ${\textrm{Im}}[{{J}}_{ijn}],~(i<j<n)$.

There are $n(n-1)/2$ phases present in $\tilde{M}$ for $n$ generations,
but all of them are not independent. $(n-1)$ of these phases can
be removed by redefining the phases of the leptons. That leaves 
$\frac{n(n-1)}{2}-n=\frac{(n-1)(n-2)}{2}=^{(n-1)}C_{2}$
independent phases in $\tilde{M}$. This is the number of Dirac phases
and may be observed in neutrino
oscillation experiments. Let us assume that some particular $n-1$
entries are made real with appropriate rephasing. We can take all
possible pair-product of these real entries. To have non-real rephasing
invariant ${\mathcal{J}}_{ijk}$, one will have to multiply pair-product
with some complex entry. For each real pair-product there correspond
only one complex entry so that there product is a complex rephasing
invariant defined as in equation \ref{mtinvariant}. So number of
all possible pair of real entries will give the number of non vanishing
rephasing measures independent of Majorana phases which is 
$^{(n-1)}C_{2}=\frac{(n-1)(n-2)}{2}$.
This number is same as the number of physical phases present in $\tilde{M}$
as it has been analyzed earlier.

In the 2-generation case there is only one CP phase which is a Majorana
phase. Which implies there should not be any non-vanishing 
${\mathcal{J}}_{ijk}$,
which is trivial to check. In the 3-generation case there is only
one Dirac CP phase, which is 
\begin{equation}
{\mathcal{J}}_{123}=\tilde{m}_{12}~\tilde{m}_{23}~
\tilde{m}_{31}=\sum_{a,b,c}\left[m_{a1}^{*}~m_{a2}~m_{b2}^{*}~
m_{b3}~m_{c3}^{*}m_{c1}\right].
\end{equation}
Thus given a neutrino mass matrix one can readily say if this mass
matrix will imply CP violation in the neutrino oscillation experiments.

In the 4-generation case there are three CP phases in the PMNS mixing
matrix and 3-Majorana phase. The independent rephasing invariants
of Dirac phases will be given as ${\mathcal{J}}_{124}$,
${\mathcal{J}}_{134}$ and ${\mathcal{J}}_{234}$ . One dependent
rephasing invariant is ${\mathcal{J}}_{123}$ which can be expressed
as \[
{\mathcal{J}}_{123}=\frac{{\mathcal{J}}_{124}{\mathcal{J}}_{234}
{\mathcal{J}}_{134}^{*}}{|\tilde{m}_{14}
\tilde{m}_{24}\tilde{m}_{34}|^{2}}.\]
 In general, these invariants satisfy 
\begin{equation}
{\mathcal{J}}_{ijk}{\mathcal{J}}_{ikl}^{*}{\mathcal{J}}_{ilj}=
|\tilde{m}_{ij}\tilde{m}_{ik}\tilde{m}_{il}|^{2}{\mathcal{J}}_{jkl} 
\end{equation}
for $n$ generations, where $i,j,k = 1,2,...,n$. 

We summarize this section by restating for $n$-generation neutrino
mass matrix without any texture zeroes, the rephasing invariants
corresponding to the Dirac phase are $J_{ijn}, ~(i<j<n)$. If there
are texture zeroes, then some of the $J_{ijk},~ (i<j<k,~ k \neq n)$ 
could also be independent. 

\section{Texture Zeroes}

In case neutrino mass matrix contains zero textures in all the columns,
it is convenient to define independent rephasing invariants in slightly
different form as, \begin{equation}
{\mathcal{R}}_{ijnn}=\lim_{|m_{kn}|\rightarrow0}~~\frac{m_{ij}m_{in}^{*}
m_{jn}^{*}m_{nn}}{|m_{in}|~|m_{jn}|~|m_{nn}|}~~~~~(\forall k~
{\textrm{for~which}}~m_{kn}=0~{\textrm{and}}~i,j\neq n)
\label{genin}
\end{equation}
 Limit has to be taken for all zero textures present in $n$th column.
Any other quartic rephasing invariant can be expressed in terms of
these independent rephasing invariants ${\mathcal{R}}_{ijnn}$ as,
\begin{equation}
{\mathcal{I}}_{ijkl}={\mathcal{R}}_{ijnn}
{\mathcal{R}}_{klnn}{\mathcal{R}}_{linn}^{*}
{\mathcal{R}}_{kjnn}^{*}\label{genexp}
\end{equation}
 We can define independent rephasing invariant measures as, 
\begin{equation}
I_{ijn}={\textrm{Im}}~[{\mathcal{R}}_{ijnn}]~~~(i,j\neq n)
\label{reinms}\end{equation}
 Advantage of defining the independent re-phasing invariants 
${\mathcal{R}}_{ijnn}$
as the limiting case is that the expressions do not become undefined
due to presence of vanishing denominators.

Let us write above expression in a different form as, \[
{\mathcal{R}}_{ijnn}=|m_{ij}|e^{i(\theta_{ij}+\theta_{nn}-
\theta_{in}-\theta_{jn})}~~~(i,j~\neq n~and~i\leq j)\]
 Where $\theta_{kn}$ is the phase present at $(k,n)$ entry of the
mass matrix. If there are some zero textures in $n$th column, then
the phase corresponding to this zero entry present in expression of
${\mathcal{R}}_{ijnn}$ must be unphysical. Let us assume that there
is a zero texture at $(1,n)$ position in neutrino mass matrix. Then
the unphysical phase $\theta_{1}n$ will appear in some of the independent
rephasing invariants as defined above in equation 25, which will allow
us to make one of the independent rephasing invariants $R_{1jnn}$
to be real eliminating corresponding CP measure. So one rephasing
invariant measure vanishes corresponding to one zero texture in the
nth column. One independent rephasing invariant (and so one CP measure)
vanishes corresponding to the zero textures present in other than
$n$th column (or $n$th row). Thus the number of independent CP measures
$N_{CP}$ for neutrino mass matrix having $p$ zero textures and $q$
zero rows for $n$ generations is given by \begin{equation}
N_{CP}=\frac{n(n-1)}{2}-p+q\label{NoInReInMe}\end{equation}
 In the same way we can study the mass-squared matrices $\tilde{M}$
and write down the number of rephasing invariant measures independent
of Majorana phases $\widetilde{N}_{CP}$ is given as \[
\widetilde{N}_{CP}=\frac{(n-1)(n-2)}{2}-r+s\]
 where $r$ is the number of zero entries in $\tilde{M}$ and $s$
is the number of those rows whose all the entries excluding diagonal
one are zero. It should be noticed that above relation of $\tilde{N_{CP}}$
is only valid if $N_{CP}$is not zero.

\section{Application to Two-zero Texture Mass Matrices}

With our present formalism, we shall now study a class of 
3-generation neutrino
mass matrices with two-zero textures, which has been listed in
ref. \cite{grim}. There are seven such mass matrices that are consistent with
present information about neutrino masses:

$$ A_1: \left(\begin{array}{ccc}
0 & 0 & X\\
0 & X & X\\
X & X & X\end{array}\right); ~~~~~~~~(2 \leftrightarrow 3)~~~~~~~~
A_2: \left(\begin{array}{ccc}
0 & X & 0\\
X & X & X\\
0 & X & X\end{array}\right); $$ $$
B_1: \left(\begin{array}{ccc}
X & X & 0\\
X & 0 & X\\
0 & X & X\end{array}\right); ~~~~~~~~(2 \leftrightarrow 3)~~~~~~~~
B_2: \left(\begin{array}{ccc}
X & 0 & X\\
0 & X & X\\
X & X & 0\end{array}\right); $$ $$
B_3: \left(\begin{array}{ccc}
X & 0 & X\\
0 & 0 & X\\
X & X & X\end{array}\right); ~~~~~~~~(2 \leftrightarrow 3)~~~~~~~~
B_4: \left(\begin{array}{ccc}
X & X & 0\\
X & X & X\\
0 & X & 0\end{array}\right); $$ $$
C: \left(\begin{array}{ccc}
X & X & X\\
X & 0 & X\\
X & X & 0\end{array}\right); $$
From our
discussions in the previous section, there can be only one
CP phase in all these cases. We shall now identify 
the rephasing invariants in all the cases. Although all these
matrices differ in phenomenology, as far as CP violation is
concerned, the interchange of the indices $(2 \leftrightarrow 3)$
will not change any discussion. So, we shall not explicitly
discuss the models $A_2, B_2, B_4$, which can be obtained by
changing the indices $(2 \leftrightarrow 3)$ from the matrices
$A_1, B_1, B_3$ respectively.

\subsubsection*{ Case $A_1$:}
There is only one non-vanishing $I_{ij}$, which is $I_{23}$.
The lepton number conserving 
rephasing invariant measure ${J}_{123}$ is given by
\begin{eqnarray}
[ J_{123}] &=& {\rm Im} \left[ (m_{31}^* m_{32})( m^*_{22} m_{23} +
m^*_{32} m_{33})( m^*_{33} m_{31})\right] \nonumber \\
&=& |m_{31}|^2 I_{23}
\end{eqnarray}
Thus there is only one Dirac CP phase in this case, which will 
contribute to the lepton number conserving processes. The same
result is valid for $A_2$.

\subsubsection*{Case $B_1$:}

In this case all the measures $I_{ij}$ are vanishing. Even
the invariants of the form ${\cal I}_{ij} $ are all vanishing.
However, there is one CP phase as discussed in the previous section.
The invariant ${\cal I}_{122133} $  is non-vanishing, which
cannot be related to to the lower invariants by
${\cal I}_{122133} = {\cal I}_{1221}\cdot {\cal I}_{3322}/|m_{22}^2 $,
since  $m_{22}=0$. The lepton number conserving invariant is
related to this invariant by
\begin{eqnarray}
[ J_{123}] &=& {\rm Im} \left[ (m^*_{11} m_{12})(
m^*_{32} m_{33})(m^*_{23} m_{21})\right] \nonumber \\
&=& {\rm Im}[{\cal I}_{122133}] .
\end{eqnarray}
Again there are no Majorana CP phase. The analysis is same for
the case $B_2$. 

\subsubsection*{Case $B_3$:}

There is only one non-vanishing CP violating measure $I_{13}$,
which is related to the lepton number conserving measure by
\begin{eqnarray}
[ J_{123}] &=& {\rm Im} \left[ (m^*_{31} m_{32})(
m^*_{32} m_{33})(m^*_{13} m_{11} + m^*_{33} m_{31})\right] \nonumber \\
&=& |m_{32}|^2 I_{13} .
\end{eqnarray}
There are no more CP phase left in addition to the one entering
in lepton number conserving processes. Replacing the indices
$(2 \leftrightarrow 3)$ we get for the case $B_4$ a similar
relation $[ J_{123}]= |m_{32}|^2 I_{12}$.

\subsubsection*{Case $C$:}

This is the most interesting case. There are no CP violating
measures of the form $I_{ij}$, although the invariant ${\cal I}_{1123}$
is non-vanishing. So, there is one CP phase in this case, as expected. 
This is related to the CP violating measure that affects lepton number
conserving processes by
\begin{eqnarray*}
{{J}}_{123} & = & {\textrm{Im}}
[(m_{11}^{*}m_{12}+m_{31}^{*}m_{32})(m_{12}^{*}m_{13})
(m_{13}^{*}m_{11}+m_{23}^{*}m_{21})]\\
 & = & |m_{12}|^{2}{\mathcal{I}}_{1123}+|m_{13}|^{2}
{\mathcal{I}}_{1123}^{*}.
\end{eqnarray*}
Although this shows that the phase is a Dirac phase, in the special
case of $m_{12}=m_{13}$, there will not be any CP violation in
the neutrino oscillation experiments. This can be verified from the
fact that for $m_{12}=m_{13}$ the third mixing angle and hence $U_{13}$
vanishes. In this case the CP violation can originate from a
Majorana phase, since $J_{123}$ vanishes even when 
Im${\cal}_{1123}$ is non-vanishing. 

Another way to understand this is to write the mass matrix in
a different basis. When $m_{12} = m_{13}$, we can write the mass
matrix $C$ as 
$$ \left(\begin{array}{ccc}
X & X & 0\\
X & X & 0\\
0 & 0 & X\end{array}\right). $$
In this case the third generation decouples from the rest and
we know that for two generation there is only a Majorana phase,
which corresponds to non-vanishing $I_{12}$ and there is no
Dirac phase, as we stated above. This is the only example
of two-zero texture mass matrices where the CP violating phase
could be a Majorana phase, but this mass matrix
is not allowed phenomenologically. 

Thus there are no phenomenologically acceptable two-zero
texture neutrino mass matrices, which has any Majorana phase. 
The only CP phase possible in any two-zero texture 3-generation
mass matrix is of Dirac type and should allow CP violation in
neutrino oscillation experiments. 

\section{Summary}

In summary, we constructed rephasing invariant measures of CP violation
with elements of the neutrino masses in the weak basis. For an $n$-generation
scenario, in the absence of any texture zeroes there are $n(n-1)/2$
independent measures of CP violation, given by \[
I_{ij}={\textrm{Im}}~[m_{ii}m_{jj}m_{ij}^{*}m_{ji}^{*}]~~~~~~~~(i<j)\]
which corresponds to $n(n-1)/2$ independent CP violating phases.
Only $(n-1)(n-2)/2$ of these phases of CP violation can contribute
to the neutrino oscillation experiments and are independent of the
Majorana phases for which the rephasing invariant measures of CP violation
can be defined as \[
{{J}}_{ijn}=\sum_{a,b,c} {\rm Im}\left[(m_{ia}^{*}~m_{ja})~(m_{jb}^{*}~m_{nb})~(m_{nc}^{*}~m_{ic})\right]~~~~~(i<j<n).\]
We then defined invariants for mass matrices with texture zeroes
and ellaborated with some examples. We studied all the phenomenologically
acceptable 3-generation two-zero
texture neutrino mass matrices. We show that there are no Majorana
phase in any of the allowed cases.


\newpage

\begin{thebibliography}{10}

\bibitem{cp1}C. Jarlskog, Phys. Rev. Lett. \textbf{55}, 1039 (1985); Z. Phys. \textbf{C
29}, 491 (1985);

\bibitem{cp2}O.W. Greenberg, Phys. Rev. \textbf{D 32} (1985) 1841; 
\\ D. Wu, Phys.
Rev. \textbf{D 33} (1986) 860; 
\\ I. Dunietz, O.W. Greenberg and D. Wu,
Phys. Rev. Lett. \textbf{55} (1985) 2935.

\bibitem{cp3}J.F. Nieves and P.B. Pal, Phys. Rev. \textbf{D 36} (1987) 315.

\bibitem{cp4}B. Kayser, Phys. Rev. \textbf{D 30} (1984) 1023.

\bibitem{cp5}J.~R.~Ellis and M.~Raidal, Nucl.\ Phys.\ B \textbf{643} (2002)
229. 

\bibitem{cp6} G.~C.~Branco, T.~Morozumi, B.~M.~Nobre and M.~N.~Rebelo,
Nucl.\ Phys.\ B \textbf{617} (2001) 475; 
\\ G.~C.~Branco, R.~Gonzalez
Felipe, F.~R.~Joaquim, I.~Masina, M.~N.~Rebelo and C.~A.~Savoy,
Phys.\ Rev.\ D \textbf{67}, 073025 (2003).

\bibitem{seesaw} P. Minkowski, Phys. Lett. \textbf{B 67}, 421 (1977); 
\\ M.~Gell-Mann,
P.~Ramond and R.~Slansky in \textit{Supergravity} (P.~van Niewenhuizen
and D.~Freedman, eds), (Amsterdam), North Holland, 1979; T.~Yanagida
in \textit{Workshop on Unified Theory and Baryon number in the Universe}
(O. Sawada and A.~Sugamoto, eds), (Japan), KEK 1979; 
\\R.N.~Mohapatra
and G.~Senjanovic, Phys.\ Rev.\ Lett. \textbf{44}, 912 (1980);
\\J. Schechter and J.W.F. Valle, Phys. Rev. \textbf{D 22}, 2227 (1980).
\bibitem{lepto1}M.~Fukugita and 
T.~Yanagida, Phys.\ Lett.\ B\textbf{174}, 45 (1986).
\bibitem{lepto2}M.A.~Luty, Phys.\ Rev.\ D\textbf{45}, 455 (1992); 
\\R.N.~Mohapatra
and X.~Zhang, Phys. Rev. D\textbf{46}, 5331 (1992); 
\\A. Acker, H.
Kikuchi, E. Ma and U. Sarkar, Phys. Rev. \textbf{D 48}, 5006 (1993);
\\M. Flanz, E.A. Paschos and U. Sarkar, Phys. Lett. \textbf{B 345},
248 (1995); \\M. Flanz, E.A. Paschos, U. Sarkar and J. Weiss, Phys.
Lett. \textbf{B 389}, 693 (1996); \\E.~Ma and U.~Sarkar, Phys.\ Rev.\ Lett.\ {}
\textbf{80}, 5716 (1998); \\T.~Hambye, E.~Ma and U.~Sarkar, Nucl.\ Phys.\ B
\textbf{602}, 23 (2001); \\G.~F.~Giudice, A.~Notari, M.~Raidal,
A.~Riotto and A.~Strumia, Nucl.\ Phys.\ B \textbf{685}, 89 (2004);
\\W.~Buchmuller, P.~Di Bari and M.~Plumacher, Nucl. Phys. {\bf B 665}, 
445 (2003); Annals Phys.\ {} \textbf{315},
305 (2005); 
\\W.~Buchmuller, R.~D.~Peccei and 
T.~Yanagida, Ann.\ Rev.\ Nucl.\ Part.\ Sci.\ {}
\textbf{55}, 311 (2005).
\bibitem{pmns}Z.~Maki, M.~Nakagawa and S.~Sakata, Prog.\ Theor.\ Phys.\ {}
\textbf{28}, 870 (1962); \\B.~Pontecorvo, Sov.\ Phys.\ JETP \textbf{7},
172 (1958) {[}Zh.\ Eksp.\ Teor.\ Fiz.\ {} \textbf{34}, 247 (1957){]};
Sov.\ Phys.\ JETP \textbf{6}, 429 (1957) {[}Zh.\ Eksp.\ Teor.\ Fiz.\ {}
\textbf{33}, 549 (1957){]}.
\bibitem{grim}P.H. Frampton, S.L. Glashow and 
D. Marfatia, Phys. Lett. \textbf{B
536}, 79 (2002); \\W. Grimus, A.S. Joshipura, L. Lavoura and M. Tanimoto,
Eur. Phys. J \textbf{C 36}, 227 (2004); \\W. Grimus, PoS (HEP 2005)
186 {[}hep-ph/0511078{]}.
\end{thebibliography}
\end{document}